\documentclass[preprintnumbers,amsmath,amssymb,nofootinbib,twocolumn,a4paper]{revtex4}

\usepackage{graphicx}
\usepackage{bm}

\newcommand{\be}{\begin{equation}}
\newcommand{\ee}{\end{equation}}
\newcommand{\bea}{\begin{eqnarray}}
\newcommand{\eea}{\end{eqnarray}}
\newcommand{\bdm}{\begin{displaymath}}
\newcommand{\edm}{\end{displaymath}}
\newcommand{\beas}{\begin{eqnarray*}}
\newcommand{\eeas}{\end{eqnarray*}}

\begin{document}

\title{Generation of Arbitrarily Non-Gaussian Fields with a Set Correlation Structure}
\author{Iain A. Brown}
\email{ibrown@astro.uio.no}
\affiliation{Institute of Theoretical Astrophysics, University of Oslo, P.O. Box 6094, N-0315 Blindern, Norway}

\date{\today}

\begin{abstract}
\noindent Non-Gaussianity in the cosmic microwave background and the large-scale structure of galaxies provides an increasingly powerful probe of the universe. I implement an algorithm to generate realisations of fields that possess an arbitrary probability distribution function and an arbitrary power spectrum and demonstrate the code with a number of examples, including the uniform distribution, the Laplace distribution, the $\chi$ and $\chi^2$ distributions, Rayleigh and Maxwell-Boltzmann distributions. The code is available at {\tt http://sourceforge.net/projects/nongaussian}.
\end{abstract}

\maketitle

\section{Introduction}
\noindent Recent Planck constraints on the local 3-point cosmic microwave background (CMB) non-Gaussianity characterised by $f_{\mathrm{NL}}$ are extremely tight \cite{Ade:2013ydc}. Rather than necessarily implying that the primordial non-Gaussianity was almost zero, this focusses attention on more complicated non-Gaussianities featuring scale-dependence, higher-order correlations or both (see \cite{Byrnes:2010ft,Mizuno:2009mv,Ade:2013ydc,Suyama:2013nva} for some examples), or characteristics of non-Gaussianity that are not dependent on a bispectrum, such as the empirical process approach in harmonic space \cite{Hansen:2002db,Hansen:2003bu} or the widespread use of Minkowski functionals (an extremely broad field; see \cite{Ducout:2012it,Ade:2013ydc} for applications to Planck). Non-Gaussianity also arises naturally in the context of networks of cosmic defects (e.g. \cite{Hindmarsh:2009qk}). At the same time, due to the nonlinearity of evolution in the later universe, non-Gaussianity is an inevitable feature of large-scale structure (LSS) surveys, both at the two-point level, where it possesses a degeneracy with the galactic bias \cite{Bruni:2011ta}, and on large-scales at the 3-point level and above. $n$-body codes initialised at late stages in the evolution of the universe may also need to take the non-Gaussianity into account, and Rayleigh distributions are employed in their initial conditions (e.g. \cite{Bardeen:1986,Sirko:2005uz}). Future tests of the CMB non-Gaussianity, and probes of upcoming high-precision LSS datasets such as the Euclid \cite{Laureijs:2011mu} and SKA \cite{2013IAUS..291..337T} projects will produce therefore rely on the generation of realised fields possessing in principle an arbitrary probability distribution function (PDF) and an arbitrary power spectrum. This is far from a trivial problem.

It is straightforward to generate a non-Gaussian field; one merely needs to populate a grid with variables chosen from a particular distribution, with the simplest example being a uniform distribution. Unfortunately this procedure presents us with no control over the correlation structure characterised by the Fourier space power spectrum, and the result is a uniform field with a white noise power spectrum. Conversely, it is extremely well known that a Gaussian field with arbitrary correlation structure can be generated by populating a grid in Fourier space with Gaussian variables of a particular power spectrum. Since a Gaussian field maps onto another Gaussian field under Fourier transformation, the result in coordinate space is a Gaussian field with the required correlation structure. Unfortunately, the simple solution of populating a field in Fourier space with variables chosen from a non-Gaussian distribution doomed, since under Fourier transformation the non-Gaussian form will be driven towards Gaussianity. The issue of the generation of non-Gaussian fields been considered with good generality by, for instance, \cite{Weinberg:1992,Contaldi:2001wr,Vio:2001cm,AvilaReese:2003ue,Rocha:2004ke,Smith:2006ud,Fergusson:2009nv}, but the combination of flexible, user-friendly techniques applicable to a wide range of PDFs and a modern implementation is missing. Otherwise it is common to enforce bispectra of the local form (as in for instance \cite{Viel:2008jj}), to employ a specific physical model that manipulates Gaussian fields (as in for instance \cite{Scoccimarro:2000qg,Liguori:2003mb}), or to focus tightly on the CMB (as in \cite{Liguori:2003mb,Rocha:2004ke,Fergusson:2009nv}). It seems timely to re-examine the issue with the intention of producing a highly flexible, publicly-available code that can generate non-Gaussian fields with arbitrary statistical nature, with a minimum of work for the user.

Outside of cosmology, this issue has attracted frequent attention and general techniques have been developed. See \cite{Grigoriu:1984,Yamazaki:1988,Deodatis:1996,Grigoriu:1998,Gurley:1998,Deodatis:2001,Masters:2003,Phoon:2005,Bocchini:2008,Shields:2011} for a representative but non-exhaustive sample of studies and \cite{Shields:2011} for a recent overview of the field. The most common approach is to employ Grigoriu's translation, or spectral distortion, method \cite{Grigoriu:1984} in which a Gaussian field is generated with a carefully chosen spectrum and the field is passed through a transformation in coordinate space to distort the statistical nature \cite{Yamazaki:1988,Deodatis:1996,Grigoriu:1998,Masters:2003,Bocchini:2008}. A closely related method is to employ a similar Hermite transformation in which the skewness and kurtosis of the field are similarly distorted \cite{Gurley:1998,Masters:2003}. Other algorithms include Karhunen-Loeve expansion (e.g. \cite{Sakamoto:2002J,Sakamoto:2002P,Phoon:2005}), which we will not pursue here.

The spectral distortion approach has been shown to be reliable and robust, although published implementations have tended to be outside of astrophysics and to focus on one-dimensional fields. This approach has the disadvantage that it is extremely slow, requiring multiple iterations before an input power spectrum is found which is suitably accurate. As discussed recently in \cite{Shields:2011} it is also not always possible to find a power spectrum that is compatible with the target PDF -- implying that it is not always possible to produce an output field with a power spectrum arbitrarily close to the target spectrum. In such cases, however, it is possible to find output power spectra that are ``close enough'', particularly given numerical noise and the strong infra-red divergences that plague realisations of the strongly red fields frequent in cosmology. Hermite transformation does not need an inverse cumulative distribution function to be defined. However, it is not very general as it only fixes the first four moments of the distribution to the required form, and it is not significantly faster than the spectral distortion method.

As a result I focus exclusively on the spectral distortion method. This has been discussed in an astrophysical and cosmological context in \cite{Vio:2001cm,Vio:2002zr}, where the authors presented pseudocode and results for some common distributions. This work does not seem to be widely appreciated in the community and I feel it is worth revisiting the method in the light of the recent focus on non-Gaussianity. Furthermore, the authors do not seem to have made their implementation public, and it was restricted to the probability distribution functions which they defined. The early work by \cite{Weinberg:1992} considered three highly non-Gaussian PDFs (positively and negatively skewed log-normal distributions and a Laplace distribution) but their method is not entirely straightforward to generalise to a specific specified input PDF.\footnote{While this can certainly be done it takes a reasonable amount of work.} Log-normal fields were likewise considered in \cite{AvilaReese:2003ue}, while in principle arbitrary PDFs were allowed in \cite{Rocha:2004ke} in the context of CMB realisations.

In this paper I implement an algorithm similar to that in \cite{Vio:2001cm}, based on the spectral distortion method of \cite{Grigoriu:1984} and most closely modelled on those in \cite{Deodatis:2001} and \cite{Bocchini:2008}. Unlike \cite{Vio:2001cm,Vio:2002zr} I restrict my attention to isotropic power spectra, with $\left<a(\mathbf{k})a^*(\mathbf{p})\right>=P(k)\delta(\mathbf{k}-\mathbf{p})$, although it would be straightforward to generalise the code to include anisotropic power spectra (useful for, for example, anisotropic models of inflation as in \cite{Pullen:2007tu}). I consider only univariate distributions -- generalisation would be similar to that in \cite{Vio:2002zr}. Previous approaches and implementations have both advantages and disadvantages, and a common disadvantage is a restriction to particular classes of PDF -- for instance, those which can be specified analytically, or for which the cumulative distribution function (CDF) or even its inverse, the quantile function (QF), can be specified analytically. In contrast, my approach emphasises flexibility at the potential cost of some processing speed.

The spectral distortion method requires only the target QF and the target power spectrum. In my implementation, should the QF be known analytically then it can be readily included in the code; in other situations, the CDF can be specified analytically and the QF found numerically, or the PDF can be specified analytically and the CDF and QF found numerically; or else the PDF itself can be input numerically. I demonstrate the code with a range of common PDFs: Gaussian, uniform, Laplace, $\chi^2$, $\chi$ and its subcases (the Rayleigh and Maxwell-Boltzmann distributions), the Planck distribution, and also implement some general distributions that include the above as subcases. I assume for simplicity that the spectra are power-law, with a hard ultra-violet cut-off at the Nyquist frequency of the grid. The resulting code, written in Fortran 95, is available at {\tt http://sourceforge.net/projects/nongaussian}, under a permissive (BSD) license.

It is important to emphasise the contribution this code makes: it is programmed in a self-contained, module that easily interfaces with both Fortran and C/C++; it is capable of taking numerically-specified input PDFs and therefore does not rely on complicated analytical forms or inaccurate semi-analytic approximations; while the first run with a new statistical nature, power spectrum and grid-size requires the iterative spectral distortion, further fields can be generated with the same recovered input spectrum (in a manner similar to the algorithms in \cite{Bocchini:2008,Shields:2011}); and the code is straightforward to extend to further analytical cases when desired. Of these perhaps the most powerful is that one does not need to specify any analytics (an issue with the implementation of \cite{Vio:2001cm} highlighted by \cite{Rocha:2004ke}, inherited from the scheme of \cite{Grigoriu:1998}) -- and that the code is available publicly and can be freely modified and redistributed. I focus on periodic, cubic grids which would be most useful for studies in later cosmology, although the algorithm can be readily adapted to CMB analysis.

The paper is structured as follows. \S\ref{Overview} presents the algorithm. In \S\ref{FullyAnalyticDistributions} it is applied to distributions for which the PDF, CDF and QF are all analytically well-defined. \S\ref{MixedDistributions} considers distributions for which either the CDF, the QF or both can only be found numerically, and in \S\ref{NumericDistributions} two fully-numerically specified PDFs are presented. Unless otherwise specified, grids are small and of size $64^3$.

\section{Overview of the Algorithm}
\label{Overview}
\noindent There are two algorithms involved in generating the non-Gaussian fields, which I refer to as the quantile transformation and the spectral distortion. The intention is to generate a Gaussian field with a power spectrum carefully chosen such that when the field is passed through a quantile transformation the resulting field has the desired spectrum. Schematically, the final procedure is as follows:
\begin{enumerate}
\item Select a target power spectrum $\mathcal{P}(k)$ and PDF $p(x)$, implying the target CDF $C(x)$ and QF $Q(x)=C^{-1}(x)$.
\item Generate a Gaussian field on a grid of size $l_{\mathrm{dim}}^3$ with the tuned spectrum $\mathcal{P}_I(k)$
\item Transform the field with $x\rightarrow C^{-1}(C_G(x))$
\end{enumerate}
Here $C_G(x)$ is the CDF of a pure Gaussian. If the tuned spectrum $\mathcal{P}_I(k)$ is chosen appropriately, the power spectrum of the resulting field will be the target spectrum $\mathcal{P}(k)$. The production of the fields therefore involves a loop across $l_{\mathrm{dim}}^3$ to generate the field, followed by an FFT and a loop across $l_\mathrm{dim}^3$ to perform the quantile transformation.\footnote{This is slightly pessimistic, since the production of real fields does not involve a full loop across $l_\mathrm{dim}^3$, but it is a reasonable first estimate.}

Finding an appropriate $\mathcal{P}_I(k)$ is an iterative spectral distortion procedure. I employ an algorithm similar to those outlined by \cite{Grigoriu:1998,Deodatis:2001}:
\begin{enumerate}
\item Generate a Gaussian field with the input spectrum $\mathcal{P}_I(k)=\mathcal{P}(k)$
\item Transform the field with the quantile transform, $x\rightarrow Q(C_G(x))$
\item Find the power spectrum of the resulting field, $\mathcal{P}_F(k)$
\item If $\mathcal{P}_F(k)\approx\mathcal{P}(k)$ output $\mathcal{P}_I(k)$ and exit
\item Otherwise, set $\mathcal{P}_{I,\mathrm{new}}(k)=(\mathcal{P}_F(k)/\mathcal{P}_I(k))^\beta\mathcal{P}(k)$ and start again
\end{enumerate}
The parameter $\beta\approx 1$ can be chosen freely to maximise convergence. This process produces a field of the requested grid size, statistical nature and power spectrum, along with the input spectrum $\mathcal{P}_I(k)$ which can be used to generate further fields without the need for this iteration. The use of $C_G(x)$ rather than the empirically-recovered $C_F(x)$ is in line with \cite{Yamazaki:1988,Grigoriu:1998}, as opposed to, for instance, \cite{Deodatis:2001}. Since the generated field is Gaussian to a high precision -- even on grids as small as $l_\mathrm{dim}=64$ -- there seems little purpose spending additional time calculating $C_F(x)$. Indeed, doing so can induce significant (and occasionally catastrophic) fluctuations in the resulting PDF. Employing the analytical expression for the Gaussian CDF (equation (\ref{GaussianEq})) yields significantly more stable results on smaller grids, and consumes less CPU time on larger grids.

The spectral distortion iteration is extremely intensive, involving (1): a loop across $l_\mathrm{dim}^3$ to generate the field; (2): an FFT, followed by a loop across $l_{\mathrm{dim}}^3$ for the quantile transformation; (3) an FFT, followed by a loop across $l_{\mathrm{dim}}^3$ to find the resulting power spectrum; (4): a loop across $l_{\mathrm{dim}}$ to test the power spectrum; and (5) a loop across $l_{\mathrm{dim}}$ to set the new power spectrum. This scales as $l_\mathrm{dim}^3$ and is extremely punishing. Storing $\mathcal{P}_I(k)$ as a function of $l_{\mathrm{dim}}$, $\mathcal{P}(k)$ and $f(x)$ is therefore extremely useful.

The target PDF should have zero mean, ensuring that the PDFs, CDFs and QFs will be centred around $x=0$. This also implies that the root-mean square of a field is equivalent to its standard deviation. To localise the statistics around the origin it is also convenient to express the PDF with the field in units of the standard deviation or, where this is unknown or ill-defined, a convenient scale parameter such that approximately $\phi\in(-20,20)$. Means and scale parameters can be readily reintroduced once a field is introduced, with $\phi\rightarrow\phi+s\phi$. During the quantile transformation the standard deviation of the output field is normalised to that of the input Gaussian field. For a numerically-specified target PDF, the mean is removed and the field axis scaled to the standard deviation recovered from numerical integration of the second moment, $\sigma^2=\int (x-\mu)^2p(x)dx$.

The field generation routine accepts a field (initialised with the {\tt initialiseField} routine) in which the {\tt statistics} flag has been set to the required value. The user can optionally also set up the the {\tt pdf} and {\tt pdfgrid} arrays with the required input PDF. If the {\tt statistics} flag is set to a value undefined in the generation routines and the {\tt pdf} routine is not initialised then the routines will look to the file {\tt \$statistics.dat} in the working directory. At present the routines generically assume that the power spectrum is power-law in nature, but this is readily generalised. If an input spectrum $\mathcal{P}_I(k)$ for the specified statistical nature, power spectrum and grid-size is known then the routines employ it in the quantile transformation; otherwise the spectral distortion iteration is entered. A small number of input spectra employed in this paper are provided. Fields are generated on a cubic grid with periodic boundary conditions, but this could be readily generalised if required.

The provided driver routine is extremely simple; it requests the user specify the statistical nature of the field, calls the field generation routine, evaluates the first four moments (mean, variance, skewness and kurtosis) of the field, evaluates the PDF and power spectrum, and then exits. It is provided to demonstrate how one initialises and generates a non-Gaussian field with the code; henceforth the user can pass the non-Gaussian field to any routine they choose. Variables are in double precision.

\section{Probability Distribution Functions}
\noindent In this section U apply the algorithm to a broad range of PDFs. Fields are generated on small grids with $l_\mathrm{dim}=64$, and the success demonstrates that the approach will work on much larger grids. This is explicitly demonstrated with uniform fields generated with $l_\mathrm{dim}=512$.

The PDF and CDF of a Gaussian field with zero mean written in units of the standard deviation are
\be
\label{GaussianEq}
\begin{array}{c}
p_G(x)=\dfrac{1}{\sqrt{2\pi}}\exp\left(-\dfrac{x^2}{2}\right), \\
C_G(x)=\dfrac{1}{2}\left(1+\mathrm{erf}\left(\dfrac{x}{\sqrt{2}}\right)\right) .
\end{array}
\ee
Test fields are generated on this grid with power spectrum $\mathcal{P}(k)=H(k-k_c)k^n$ and $n\in\{0,-2.9\}$. The former is a white-noise field, while the latter is reasonably close to inflationary scale-invariance. $H(x)$ is the Heaviside function and $k_c$ a small-scale cutoff taken to lie at the grid's Nyquist frequency, $k_c=l_\mathrm{dim}/2$. The PDF and CDF of the input fields are presented in the background of Figure \ref{FullyAnalyticPDFs}, and the power spectra are in the leftmost panel of Figure \ref{FullyAnalyticSpectra}. Note in particular a damping on large scales, associated with the inevitable infra-red cut-off that arises on a finite grid. A slice through the realisation is presented in the centre of Figure \ref{Realisations}.\footnote{Note that this realisation is generated on a low-resolution grid; the straight edges in the plot are unphysical.} For illustrative purposes I recover the input spectra $\mathcal{P}_I(k)$ from $n\sim 10$ iterations, and then generate two new Gaussian fields with the same random seed to pass through the quantile transformation, which allows a direct comparison between the raw Gaussian and the resulting fields.

The input PDF is employed with zero mean and fields are output with standard deviation normalised to that of the input Gaussian. Both the mean and the standard deviation can easily be reintroduced if so desired, as can scale parameters in the PDF. These details are left to the calling routines to preserve the generality of the production routine. Output PDFs and CDFs are presented in units of the field's standard deviation.

The quantile transformation in greater detail is
\begin{enumerate}
\item At each point $\{i,j,k\}$ in the grid, find the Gaussian field value $\phi(i,j,k)$
\item Find the CDF for this field value, $C_G(\phi)$
\item Transform the field with $\phi\rightarrow Q(C_G(\phi))$
\item Scale the resulting field to a desired value
\end{enumerate}
Three approaches to the desired output probability distribution functions are implemented: fully analytic, mixed analytic/numerical, and fully numerical approach.

In the fully analytic approach, both the PDF and QF of the target distribution are specified analytically, and the CDF is not required. In a mixed approach, the PDF is known analytically while the QF is not (or is sufficiently complicated to make a numeric approach preferable). In a fully numeric approach the PDF itself is specified with an input array. Whenever the cumulative distribution function is known analytically it is employed analytically; however, I demonstrate (with Laplace fields) that the error in directly recovering the CDF by integrating across the PDF is negligible.

Where the QF is not known analytically, as is typically the case, it is recovered from an inverse interpolation: the CDF is set up in a table $C_i$ with abscissas $x_i$, and recovered with an interpolation over the abscissas, $C(x)=I(x_i,C_i|x)$. The QF is found with an inverse lookup, $Q(x)=I(C_i,x_i|x)$. This simple approach is fast but relies on a CDF sampled sufficiently densely. In practice, for $C(x)\in[0,0.997)$ is sampled with 4,000 points and the region $C(x)\in(0.997,1]$ with tens of thousands of points. This ensures that the transformation remains stable for the input PDFs considered thus far but it may be necessary to modify the sampling for particularly problematic PDFs.

This table illustrates the PDFs implemented natively in the code. Explicit subcases will be used where these are defined; additional subcases can be explicitly implemented with relative ease.\footnote{General properties are defined in the {\tt initialisePDF} subroutine, while the analytic PDF, CDF and QF are respectively defined in the {\tt targetPDF}, {\tt targetCDF} and {\tt targetQF} functions.} In the following table, items in square brackets denote those implemented analytically in the code. Quantities in brackets are the parameters of the distributions; generically, $d$ denotes the number of degrees of freedom, $p$ the characteristic power index, and $s$ a scale parameter. For convenience, the input PDF will be written with the field in units of the scale parameter or the standard deviation where appropriate.
\begin{itemize}
\item Log-logistic$_{(\alpha,\beta)}$ [P, C, Q]
\item Planck$_{(T)}$ [P]
\item Generalised gamma, $\tilde{\gamma}_{(d,p,s)}$ [P]
  \begin{itemize}
  \item Weibull$_{(d,s)}=\tilde{\gamma}_{(d,d,s)}$
  \item $\chi_{(d)}=\tilde{\gamma}_{(d,2,2)}$ [P, C]
    \begin{itemize}
    \item Rayleigh=$\tilde{\gamma}_{(2,2,2)}$ [P, C]
    \item Maxwell-Boltzmann=$\tilde{\gamma}_{(3,2,2)}$ [P, C]
    \end{itemize}
  \item Gamma, $\gamma_{(d,s)}=\tilde{\gamma}_{(d,1,s)}$ [P, C]
    \begin{itemize}
    \item $\chi^2_{(d)}=\tilde{\gamma}_{(d,1,2)}$ [P, C]
    \item Exponential$_{(\lambda)}=\tilde{\gamma}_{(1,1,\lambda^{-1})}$ [P, C, Q]
	\item Erlang$_{(d,s)}=\tilde{\gamma}_{(d\in \mathbb{Z}^+,1,s)}$
    \end{itemize}
  \item Nakagami$_{m,\Omega}=\tilde{\gamma}_{(2m,2,\sqrt{\Omega/m})}$
  \end{itemize}
\item Generalised error distribution, $\tilde{E}_{(p,s)}$ [P, C]
  \begin{itemize}
  \item Gaussian$_s=\tilde{E}_{(2,s)}$ [P, C]
  \item Laplace$_s=\tilde{E}_{(1,s)}$ [P, C, Q]
  \item Uniform$_\sigma=\tilde{E}_{(p\rightarrow\infty,\sigma)}$ [P, C, Q]
  \end{itemize}
\end{itemize}

The following sections consider each form of distribution, selecting subcases of particular interest. These can be taken as archetypes for the fully analytic, mixed and fully numeric classes of distribution. Note again, however, that \emph{any} PDF can be employed; one merely needs to specify the PDF in an input table. This is demonstrated with a uniform distribution and a convoluted distribution in \S\ref{NumericDistributions}.

\subsection{Fully Analytic Distributions}
\label{FullyAnalyticDistributions}
\noindent A fully analytic approach requires that both the target PDF and QF can be specified. These are hard-coded in, while any parameters (if relevant) can be passed into the code, or are requested at the start of the process. For these fields knowledge of the target CDF is not required, although they are presented to confirm the output CDFs.

\vspace{0.5em}\noindent\textbf{Uniform Distributions}\newline
\noindent In principle, generating a uniform field is easy: populate it in coordinate space with numbers drawn from a uniform distribution. However, the correlation structure of this field will be white-noise and power spread across the entire grid, which is certainly not ideal if one wishes to manipulate the field any further. Running through quantile transformation provides full control over the field.

\begin{figure*}
\includegraphics[width=\textwidth]{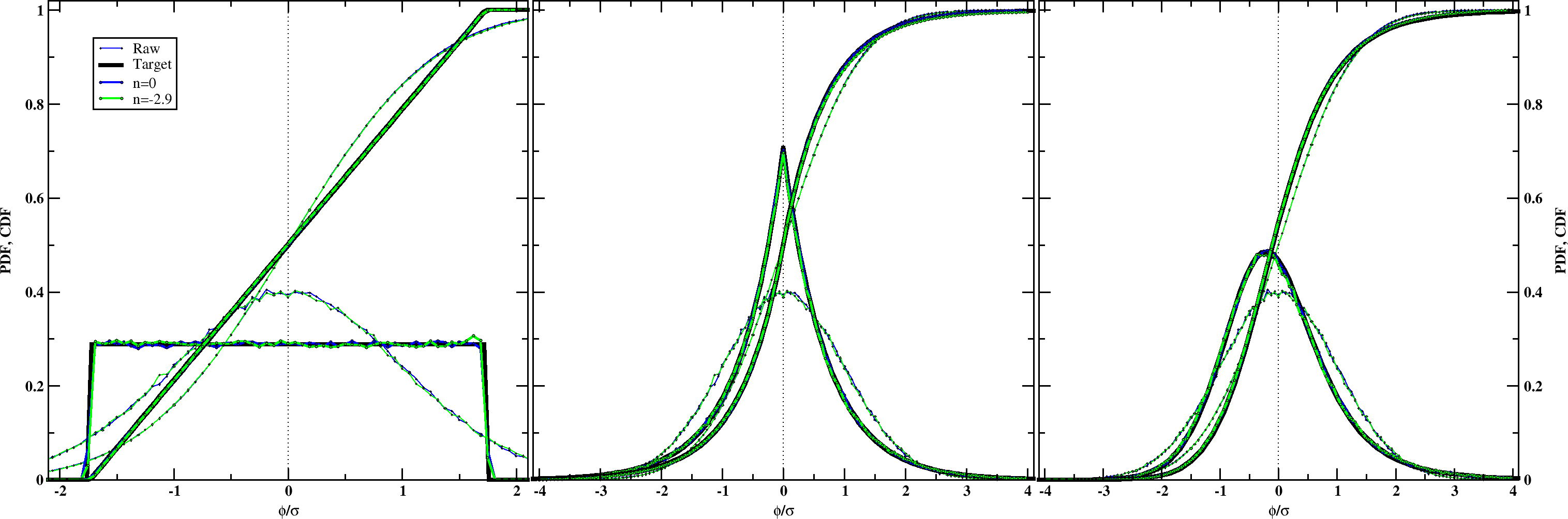}
\caption{PDFs and CDFs for the uniform distribution (left), Laplace distribution (middle) and log-logistic distribution with $\{s,p\}=\{1,9\}$ (right). The raw input Gaussian fields are plotted in the background.}
\label{FullyAnalyticPDFs}
\end{figure*}

This distribution is fully specified and in general takes two parameters, $a$ and $b$, specifying the ends of the distribution. The mean is $\mu_1=(a+b)/2$ and the variance $\sigma^2=(1/2)(b-a)^2$. Enforcing a zero mean reduces these to a single parameter, $a$. Expressing the field in units of the standard deviation is equivalent to setting $a=\sqrt{3}$. The PDF, CDF and QF are then
\be
\begin{array}{c}
\vspace{0.5em}
p(x)=\left\{\begin{array}{rl} \dfrac{1}{2\sqrt{3}}&, \quad x\in\left[-\sqrt{3},\sqrt{3}\right],\\0&, \quad\mathrm{otherwise}\end{array}\right., \\
\vspace{0.5em}
C(x)=\left\{\begin{array}{rl} 0&,\quad x<-\sqrt{3}, \\ \dfrac{1}{2}\left(\dfrac{x}{\sqrt{3}}+1\right)&,\quad x\in[-\sqrt{3},\sqrt{3}], \\ 1&,\quad x\geq\sqrt{3}\end{array}\right., \\ \quad
Q(x)=\sqrt{3}(2x-1), \quad x\in(0,1) .
\end{array}
\ee
The PDFs and CDFs of the generated fields are presented in Figure \ref{FullyAnalyticPDFs}. Note that due to the form of the distribution, power will only extend to $\phi=\pm\sqrt{3}\sigma$. The target statistics are drawn in black while the output are presented in points, with the input Gaussian in the background.

\begin{figure*}
\includegraphics[width=\textwidth]{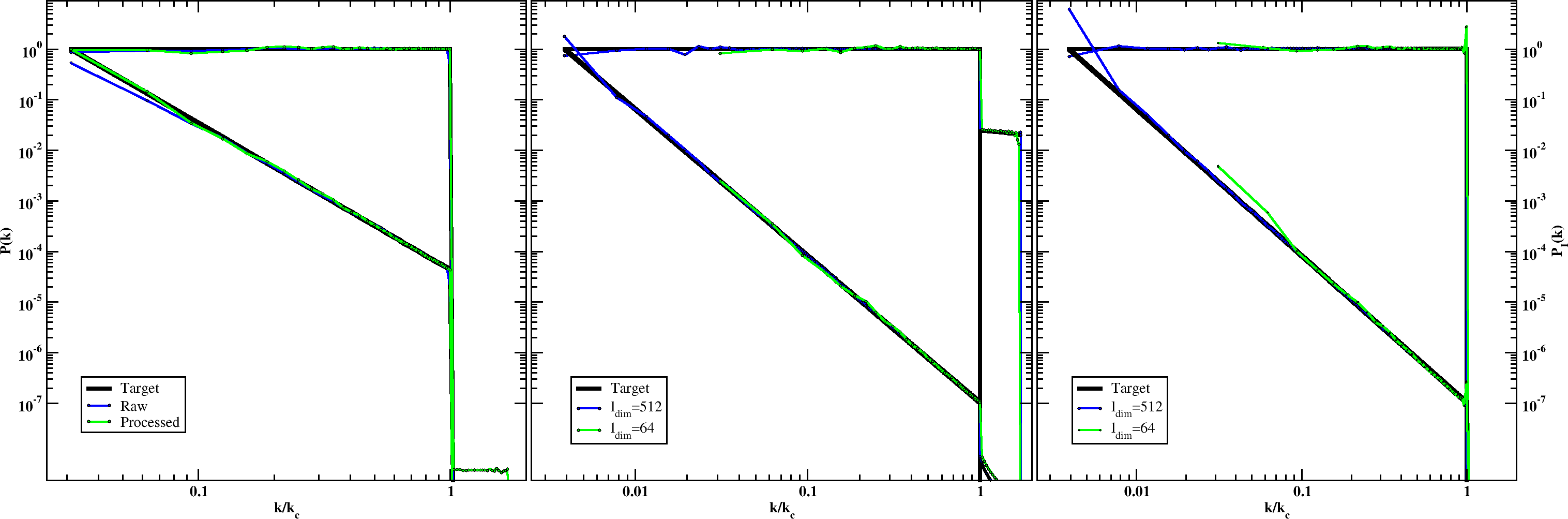}
\caption{Left: raw input and processed spectra for Gaussian fields. Middle: power spectra for white-noise and near-scale-invariant uniform fields on small and large grids. Right: Input spectra $\mathcal{P}_I(k)$ for the uniform fields.}
\label{FullyAnalyticSpectra}
\end{figure*}

By construction the output power spectra are extremely similar for each field, so I will not present them for every distribution function. However, it is instructive to compare the spectra found for small and large grids, and these are shown for the uniform distribution in Figure \ref{FullyAnalyticSpectra}. In the middle panel are the output power spectra, while in the right panel are the input spectra $\mathcal{P}_I(k)$ that generate these. There are two important, related features of the output spectra: the damping on large scales has been corrected by the spectral distortion, which is clearly apparent in the input spectra; less pleasing is an excess of power for $k>k_c$.

There are two possible causes of this. The first is a result of the incompatibility of the chosen power spectra and the chosen statistics while employing a quantile transformation, as discussed in \cite{Shields:2011}. However, for power law fields the dominant contribution instead arises from the spectral distortion itself. Spectral distortion corrects the large-scale damping at the expense of a leakage of power onto smaller scales. The small-scale tail is one or two orders of magnitude lower than the rest of the power spectrum and the error induced by these tails will be negligible, particularly for near-scale-invariant fields. It seems unlikely for practical purposes that the leakage of power would be significant.

If aliasing becomes problematic -- such as if fields need to be cubed, or multiple Laplacians applied -- then the issue can be significantly lessened by preventing the spectral distortion from operating on large scales.\footnote{It is also possible to remove the modes with $k>k_c$ immediately after the quantile transformation and before the spectral distortion, with a consequent distortion of the output PDF that tends to soften the central peaks. For some distributions closer to Gaussian the distortions may be acceptable but for distributions further from Gaussian -- such as the uniform -- the distortions are significant. In particular, removing the small-scale modes for a uniform field turns the sharp-edged PDF into a loose hump.}

It might be worried that while the additional power on large scales improves the two-point statistics it will induce large-scale errors in the higher-order statistics. It seems that this is not the case. The fields are generated on a finite grid, and the large-scale damping arises due to the scarcity of large-scale modes that can satisfy statistical homogeneity. An increasing lack of such modes for higher-order corrections means that at worst they will be unaffected, once possible contributions from the two-point moment have been subtracted. So long as the effects of aliasing when manipulating the fields are negligible, the results are more accurate, rather than less.

\begin{figure*}
\includegraphics[width=\textwidth]{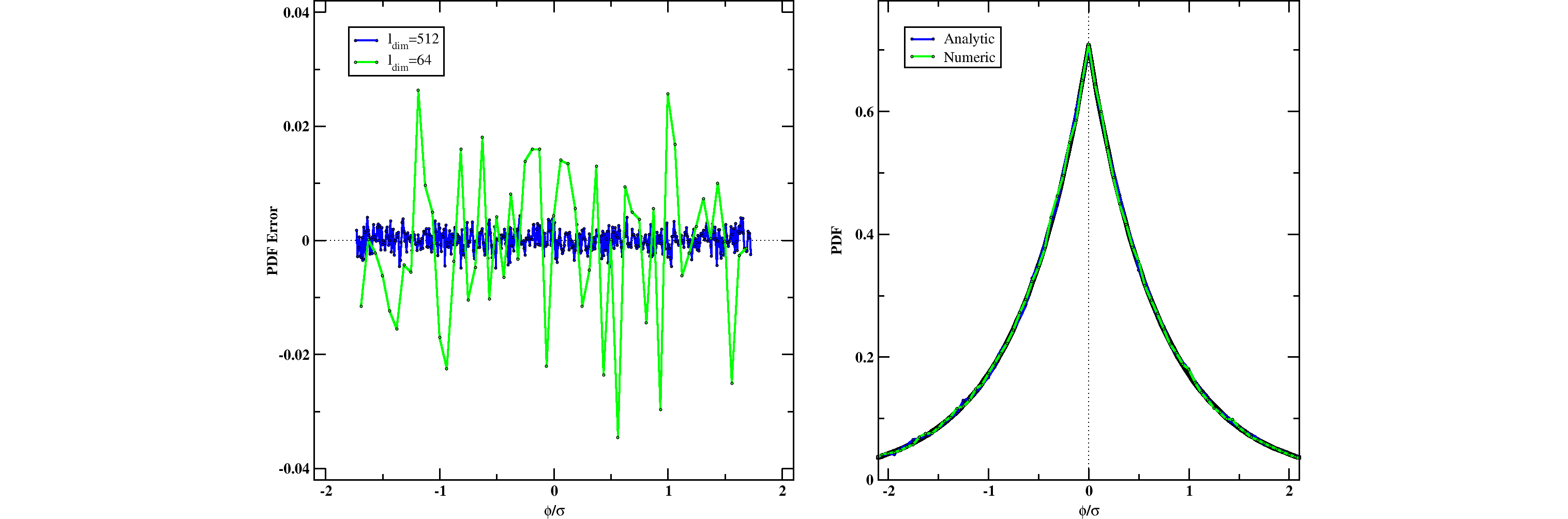}
\caption{Left: error in the PDFs for a white-noise uniform field on large and small grids. Right: White-noise Laplace fields generated with analytic and numeric quantile functions.}
\label{AnalyticComparisons}
\end{figure*}

The left panel of Figure \ref{AnalyticComparisons} shows the fractional error $E(x)=(p_\mathrm{Output}(x)-p_\mathrm{Target}(x))/p_\mathrm{Target}(x)$ for the PDFs of white-noise fields generated on grids with $l_\mathrm{dim}=64$ and $l_\mathrm{dim}=512$, with an obvious improvement in the accuracy of the transformation. However, even for the small grids the error is $\lesssim 4\%$. For other distributions with power in a lengthy tail numerical noise becomes increasingly significant and errors can be $\sim 30\%$ although the PDF is accurately recovered in regions of high probability. The noise level reduces when averaging across multiple realisations.

A uniform distribution has a zero skewness and an excess kurtosis $\gamma_2=-6/5$. Estimates from only two white-noise fields with $l_\mathrm{dim}$ with their standard deviations are $\gamma_1=2.02\left(1\pm 1.88\right)\times 10^{-3}$, consistent with zero, and $\gamma_2=-(6/5)\left(1\pm 1.8\times 10^{-5}\right)$ -- an error of two parts in ten thousand.

The white-noise uniform field is plotted in the left panel of Figure \ref{Realisations}. By eye the difference in clustering between the uniform and Gaussian fields is not immediately obvious although one can convince oneself that there is a slight difference; more obvious are the hot- and cold-spots present in the Gaussian field that are absent in the uniform, due to the Gaussian's extended tails.

\vspace{0.5em}\noindent\textbf{Laplace Distributions}\newline
\noindent The Laplace distribution describes the distribution of the difference between two exponentially-distributed variables, and is a one-parameter family. Laplace distributions were considered in \cite{Weinberg:1992}. In terms of this parameter $s$ the variance is $2s^2$. Expressing the field in terms of the standard deviation the PDF, CDF and QF are
\be
\begin{array}{c}
\vspace{0.5em}
p(x)=\sqrt{2}\exp(-\sqrt{2}x),\\
\vspace{0.5em}
C(x)=\left\{\begin{array}{rl}\dfrac{1}{2}\exp(\sqrt{2}x)&, \quad x<0\\1-\dfrac{1}{2}\exp(-\sqrt{2}x)&,\quad x>=0\end{array}\right., \\
Q(x)=-\dfrac{1}{\sqrt{2}}\mathrm{sgn}\left(x-\dfrac{1}{2}\right)\left(1-\exp\left(-\sqrt{2}\left|x\right|\right)\right) .
\end{array}
\ee
The PDFs and CDFs are plotted in Figure \ref{FullyAnalyticPDFs}. A Laplace field has a vanishing skewness and an excess kurtosis $\gamma_2=3$. The mean and standard deviation of two white-noise fields with $l_\mathrm{dim}=64$ give $\gamma_1=0.4222\left(1\pm 2.438\right)\times 10^{-2}$, consistent with zero, and $\gamma_2=3\left(1.001\pm 0.081\right)$.

The Laplace distribution permits a simple test of the numerical recovery of the QF. The right panel of Figure \ref{AnalyticComparisons} shows the PDF of Laplace fields generated with an analytic CDF and quantile, and with a quantile recovered by inverting a numerically-integrated CDF. The differences are practically indistinguishable by eye, while the fractional errors are quantitatively similar (albeit relatively large in comparison to the uniform case, due to numerical noise in the poorly-sampled extended tails) in both cases.

\vspace{0.5em}\noindent\textbf{Log-Logistic Distributions}\newline
\noindent The log-logistic provides a useful example of a fully-analytic distribution, broadly similar to the log-normal distribution, but which is not a subcase of the generalised $\gamma$ and error distributions. It is employed for instance in situations involving survival rates. In contrast to the previous cases, the log-logistic distribution is restricted to the positive real line. This is a two-parameter family which can be expressed in terms of scale and power parameters $\{s,p\}$. The PDF, CDF and QF are
\be
\begin{array}{c}
\vspace{0.5em}
p(x)=\dfrac{p}{s}\dfrac{(x/s)^{p-1}}{\left(1+(x/s)^p\right)^2}, \\
\vspace{0.5em}
C(x)=\dfrac{x^p}{s^p+x^p}, \quad
Q(x)=s\left(\dfrac{x}{1-x}\right)^{1/p} .
\end{array}
\ee
Both $p$ and $s$ are positive-definite. The mean and variance of this distribution can be found to be
\be
\begin{array}{c}
\mu_1=\dfrac{s}{\mathrm{sinc}(\pi/p)}, \\
\sigma^2=s^2\left(\dfrac{1}{\mathrm{sinc}(2\pi/p)}-\dfrac{1}{\mathrm{sinc}^2(\pi/p)}\right) .
\end{array}
\ee
For $\mu_1$ to be defined requires $p>1$, while for $\sigma$ to be defined requires $p>2$. To avoid issues likely to arise with distributions with ill-defined means and variances, $s>1$ and $p>2$ are imposed at all times. The pattern continues; for the $n$th central moment to exist requires $p>n$. For a distribution with $p=3$, for instance, the mean and standard deviation are well defined but the skewness is not. This is reflected in the large errors that occur when $n\lesssim p$.

Due to the form of the variance, it is more straightforward to generate the field directly and rescale it. In the right panel of Figure \ref{FullyAnalyticPDFs} are the PDF and CDF of a log-logistic field with $s=1$ and $p=9$. The PDF is presented with the field in units of the standard deviation and the input PDF has been rescaled appropriately. In the left panel of Figure \ref{PDFRanges} are log-logistic fields with $\{s,p\}$=$\{1,3\}$, $\{1,9\}$ and $\{2,5\}$. The evaluated skewnesses and excess kurtoses, from the average of ten fields at $l_\mathrm{dim}=64$ and their standard deviation, are
\bdm
\begin{array}{r|cc}
\{s,p\} & \gamma_1 & \gamma_2 \\
\hline
1,3 & 9.369\pm 0.079 & 353.8\pm 51.6 \\
1,9 & 1.060\left(0.996\pm 0.017\right) & 4.215\left(0.986\pm 0.073\right) \\
2,5 & 2.485\left(0.983\pm 0.0599\right) & 26.56\left(0.808\pm 0.277\right)
\end{array}
\edm
where they have been expressed in the form $\gamma_{\mathrm{Expected}}\left(a\pm\sigma_a\right)$. Note that $\gamma_1$ and $\gamma_2$ are undefined in the first instance, and that $\gamma_2$ is an edge case in the last, reflected in the large errors on the kurtosis.

\subsection{Mixed Analytic/Numeric Distributions}
\label{MixedDistributions}
\noindent In the mixed cases, the PDF is specified analytically and the QF recovered numerically. The CDF can be specified in either manner depending on convenience.

\vspace{0.5em}\noindent\textbf{Gaussian Distribution}\newline
\noindent Running Gaussian fields through the quantile transformation seems somewhat superfluous, but there may be occasions on which it will be useful. The quantile transformation pushes a noisy low-resolution PDF closer to Gaussian, while the spectral distortion corrects the punishing large-scale damping. The PDF, CDF and QF of a Gaussian field with zero mean and unit variance are 
\be
\begin{array}{c}
\vspace{0.5em}
p(x)=\dfrac{1}{\sqrt{2\pi}}\exp\left(-\dfrac{x^2}{2}\right), \\
\vspace{0.5em}
C(x)=\dfrac{1}{2}\left(1+\mathrm{erf}\left(\dfrac{x}{\sqrt{2}}\right)\right).
\end{array}
\ee
The QF is difficult to calculate analytically and is recovered numerically.

In practice, the correction to the PDF, while noticeable by eye, is not statistically significant. The resulting power spectra for the processed Gaussian fields are presented in Figure \ref{FullyAnalyticSpectra}. The large-scale damping is not exhibited. More significantly, due to the close similarity of the input and output CDFs, the leakage of power onto small scales is entirely insignificant. Since it is common to study physical processes in cosmology by generating and manipulating Gaussian fields, and since these fields typically exhibit infra-red cut-offs which contaminate results, running a Gaussian field through such a ``quantile correction'' seems a natural step.

The skewness of the normal distribution and, by definition, its excess kurtosis, are zero. For completeness, the mean and standard deviation of the skewness and excess kurtosis from the two small-grid realisations of quantile-corrected Gaussian fields are $\gamma_1=\left(2.544\pm 7.290\right)\times 10^{-3}$ and $\gamma_2=\left(0.787\pm 9.775\right)\times 10^{-3}$.

\vspace{0.5em}\noindent\textbf{$\chi^2$ Distributions}\newline
\noindent $\chi^2$ fields are generated from the sum of squared Gaussian fields; a $\chi^2$ field with $d$ degrees of freedom is defined as
\be
x^{(d)}_{\chi^2}=\sum_{i=1}^d x_{Gi}^2 .
\ee
Such fields are familiar in cosmology from electromagnetism, where the energy density and isotropic pressure of Gaussian electromagnetic fields will be $\chi^2$ in form. Certainly $\chi^2$ fields can be generated by producing multiple Gaussian fields and summing their squares as in \cite{Scoccimarro:2000qg}, but it can be preferable to generate them directly for two reasons: speed and flexibility. Not only is generating a Gaussian realisation itself relatively slow, each then must be transferred from Fourier into coordinate space, then squared, and then summed -- and each step is computationally punishing. Furthermore, the direct approach provides no control over the correlation nature of the resultant fields, which will consequently resemble those evaluated in for instance \cite{Brown:2005kr,Paoletti:2008ck} with an extended and characteristic decay. Worse, to avoid aliasing, the Gaussian fields must be generated with a cut-off at \emph{half} the Nyquist frequency, which limits the dynamic range dramatically. The quantile transformation provides full control over the power spectrum.

\begin{figure*}
\includegraphics[width=\textwidth]{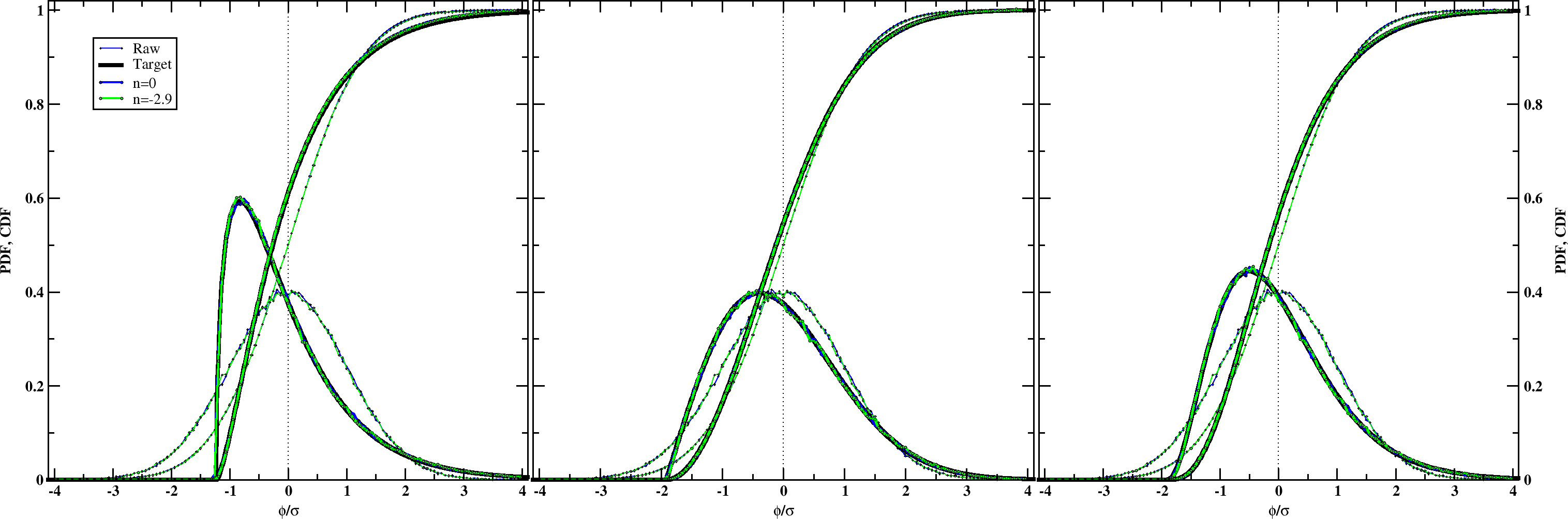}
\caption{PDFs and CDFs of $\chi^2$ fields with three degrees of freedom (left), $\chi$ fields with two degrees of freedom (Rayleigh fields; middle) and Planck fields (right).}
\label{MixedPDFs}
\end{figure*}

\begin{figure*}
\includegraphics[width=\textwidth]{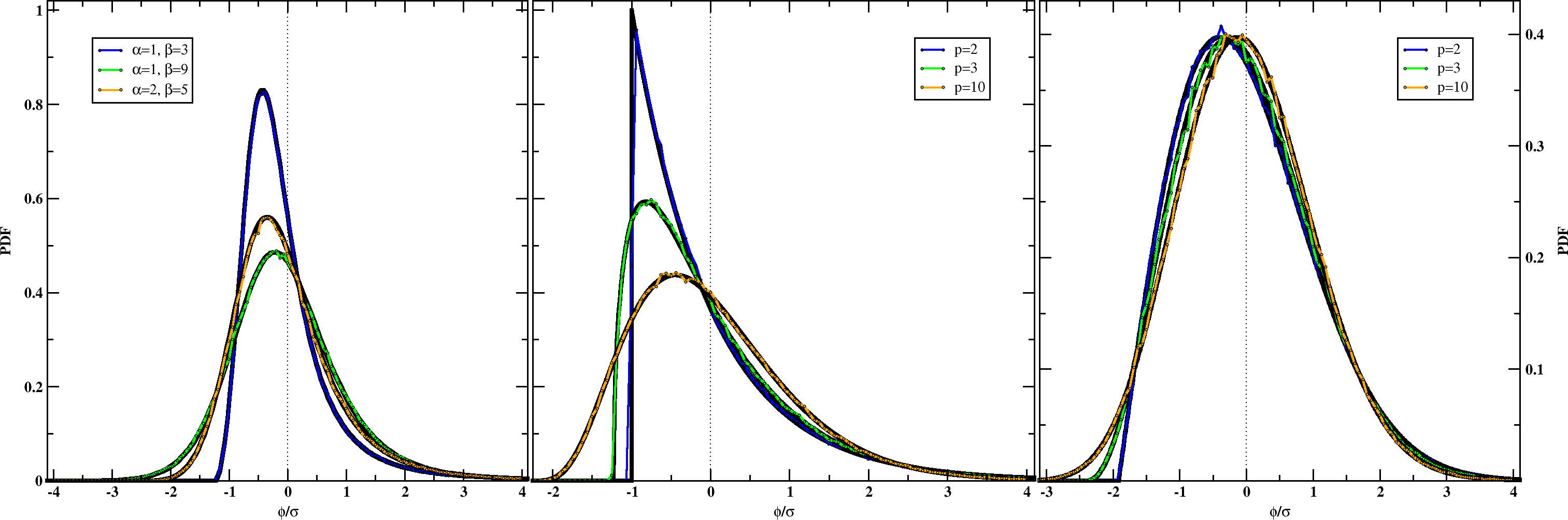}
\caption{PDFs of log-logistic fields (left), $\chi^2$ (middle) and $\chi$ (right) for various parameters.}
\label{PDFRanges}
\end{figure*}

Fields with $d$ degrees of freedom have the PDF and CDF
\be
\begin{array}{c}
\vspace{0.5em}
p(x)=\dfrac{1}{2^{d/2}\Gamma\left(d/2\right)}x^{d/2-1}\exp\left(-\dfrac{x}{2}\right), \\
C(x)=P\left(\dfrac{d}{2},\dfrac{x}{2}\right), \\
\end{array}
\ee
where $P(a,x)=\gamma(a,x)/\Gamma(a)$ is the regularised incomplete gamma function and $\gamma(a,x)$ the lower incomplete gamma function. Efficient routines exist to evaluate this with arbitrary accuracy, so only the QF must be recovered numerically, although it proves convenient to evaluate the CDF by integration. This PDF (and its implied functions) are defined on the positive real line. The mean of the distribution is $r$, and its variance is $2r$; the mean is removed and the PDF in the code expressed in units of the standard deviation. The PDF and CDF of a field with three degrees of freedom are presented in Figure \ref{MixedPDFs}, and the PDFs of fields with 2, three and ten degrees of freedom in Figure \ref{PDFRanges}. The skewness of a $\chi^2$ field is $\gamma_1=\sqrt{8/d}$ while the (excess) kurtosis is $\gamma_2=12/d$. The values and their standard deviations recovered from the average of two fields with $l_\mathrm{dim}=64$ are
\bdm
\begin{array}{r|cc}
d & \gamma_1 & \gamma_2 \\
\hline
2 & 2\left(1.001\pm 0.007\right) & 6\left(1.003\pm 0.0312\right) \\
3 & \sqrt{8/3}\left(1.001\pm 0.0049\right) & 4\left(1.0029\pm 0.0290\right) \\
10 & \sqrt{8/10}\left(1.0028\pm 0.0009\right) & 1.2\left(1.0047\pm 0.0273\right)
\end{array}
\edm

\begin{figure*}
\includegraphics[width=\textwidth]{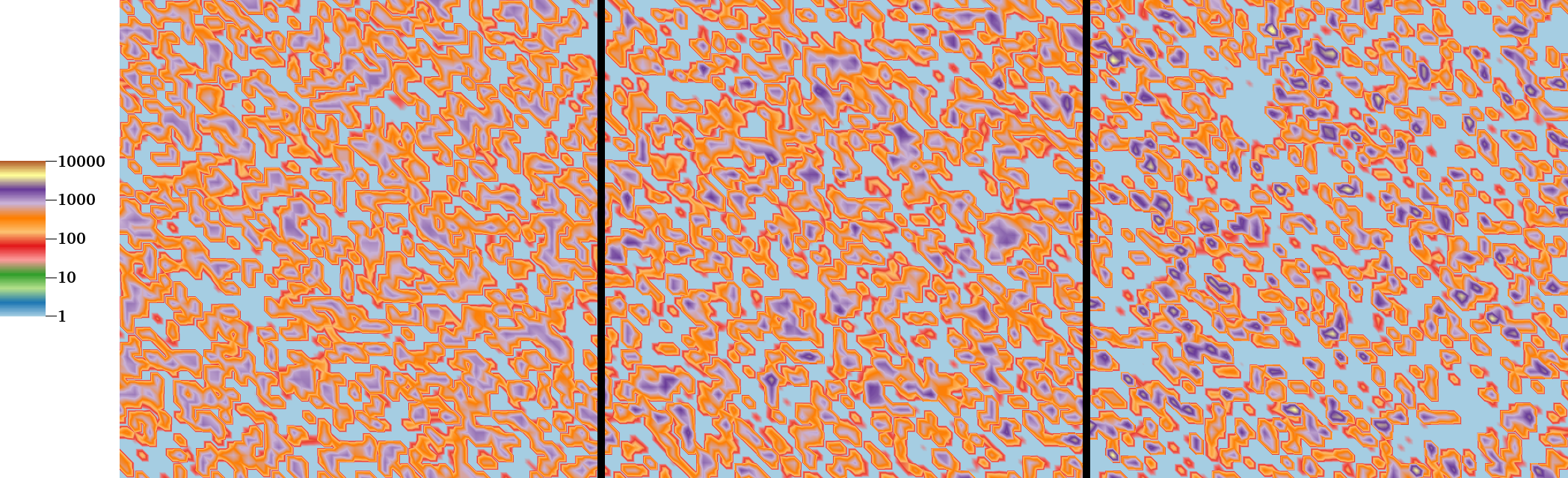}
\caption{Realisations on a small grid of uniform fields (left), Gaussian fields (middle) and Rayleigh fields (right).}
\label{Realisations}
\end{figure*}

\vspace{0.5em}\noindent\textbf{$\chi$ Distributions}\newline
\noindent The $\chi$ distribution is closely related to the $\chi^2$ distribution, and contains the well-known Rayleigh and Maxwell-Boltzmann distributions as subcases. $\chi$-distributed fields with $d$ degrees of freedom are defined by
\be
x^{(d)}_\chi=\sqrt{x^{(d)}_{\chi^2}}=\sqrt{\sum_{i=1}^d x_{Gi}^2} .
\ee
The PDF and CDF of the $\chi$ distribution are
\be
\begin{array}{c}
\vspace{0.5em}
p(x)=\dfrac{2^{1-d/2}}{\Gamma(d/2)}x^{d-1}\exp\left(-\dfrac{x^2}{2}\right), \\
C(x)=P\left(\dfrac{d}{2},\dfrac{x^2}{2}\right)
\end{array}
\ee
and the Rayleigh and Maxwell-Boltzmann distributions are found for two and three degrees of freedom respectively.\footnote{Note that in both of these cases there is an additional scaling factor in the distributions which is removed with a suitable rescaling of the field variable.} The mean and variance of this distribution are
\be
\mu_1=\frac{\sqrt{2}\Gamma\left(d/2+1/2\right)}{\Gamma(d/2)}, \quad
\sigma^2=d-\mu^2
\ee
and are removed with an appropriate translation and rescaling. The PDF and CDF of a Rayleigh field are shown in the middle panel of Figure \ref{MixedPDFs}, while \ref{PDFRanges} also shows the PDFs of a Maxwell-Boltzmann field and a $\chi$ with ten degrees of freedom. A slice through the realisation of the Rayleigh field is in Figure \ref{Realisations} and the differences in the distribution are clearly visible by eye.

Closed analytical forms for the skewness and kurtosis of a $\chi$ field are not particularly instructive. The recovered values for two realised fields, again on a small grid, are
\bdm
\begin{array}{r|cc}
d & \gamma_1 & \gamma_2 \\
\hline
2 & 0.6311\left(1.004\pm 0.006\right) & 0.2451\left(0.9891\pm 0.0389\right), \\
3 & 0.4857\left(1.005\pm 0.009\right) & 0.1082\left(1.023\pm 0.0752\right), \\
10 & 0.2374\left(1.011\pm 0.0250\right) & 8.520\left(1.200\pm 0.9457\right)\times 10^{-3} .
\end{array}
\edm

\vspace{0.5em}\noindent\textbf{The Planck Distribution}\newline
\noindent The Planck distribution describes the intensity of radiation from a black body but can also describe a PDF unrelated to those considered so far. The scale factor is typically interpreted as a temperature $T$. In units of this temperature (with $k_B=\hbar=1$) the distribution contains no free parameters and is given by
\be
p(x)=\frac{15}{\pi^4}\frac{x^3}{\exp(x)-1} .
\ee
The mean and standard deviation are
\be
\mu_1=\frac{360\zeta(5)}{\pi^4}, \quad \sigma^2=\frac{40}{21}\pi^2-\mu_1^2
\ee
These constants are approximately $\mu_1\approx 3.8322$, $\sigma\approx 2.0281$. The CDF can be found analytically but it is easier to employ numerical CDF and QF. The resulting PDFs and CDFs are in Figure \ref{MixedPDFs}. The skewness and excess kurtosis of this distribution are $\gamma_1\approx 0.9865$ and $\gamma_2\approx 1.433$. The values recovered from two realisations on the small grid are $\gamma_1=0.9865\left(1.002\pm 0.0001\right)$ and $\gamma_2=1.433\left(1.001\pm 0.027\right)$.

\vspace{0.5em}\noindent\textbf{The Log-Normal Distribution}\newline
\noindent The log-normal distribution is employed in cosmology in the generation of mock catalogues and is one of the most commonly-encountered distributions in broader fields. The standard log-normal distribution (with zero location and unit scale parameters) with shape $s$ has the PDF and CDF
\be
\begin{array}{c}
\vspace{0.5em}
p(x)=\dfrac{1}{\sqrt{2\pi}sx}\exp\left(-\dfrac{(\ln x)^2}{2s^2}\right), \\
C(x)=\dfrac{1}{2}\left(1+\mathrm{erf}\left(\dfrac{\ln x}{\sqrt{2}s}\right)\right) .
\end{array}
\ee
The standard deviation of this distribution is $\sigma=\sqrt{\exp(s^2)\left(\exp(s^2)-1\right)}$. If the field is written in units of the standard deviation then the mean is $\mu_1=\exp(-s^2/2)$, which is subtracted by a translation along the field axis. This distribution grows increasingly peaked as $s\rightarrow 1$ and is extremely tightly peaked for $s>1$. While one can certainly still generate fields with such a shape parameter it is recommended to rebin the PDF, or to evaluate it in units other than the standard deviation.

Figure \ref{Numerics} shows the PDF of a log-normal field with $s=1/2$. This has the skewness $\gamma_1=1.750\left(1.000\pm 0.011\right)$ and excess kurtosis $\gamma_2=5.898\left(0.994\pm 0.042\right)$.

\vspace{0.5em}\noindent\textbf{Generalised Distributions}\newline
\noindent Generalised distribution functions can encompass many of the previous subcases, and many are possible. For practical purposes I have implemented three: the generalised $\gamma$ distribution, the generalised error distribution, and a skew-normal distributions.

The PDFs of the $\chi$ and $\chi^2$ distributions contain a regularised incomplete gamma function. A natural extension is to define the CDF and the implied PDF
\be
\begin{array}{c}
C(x)=P\left(\dfrac{d}{p},\left(\dfrac{x}{s}\right)^p\right), \\
\Rightarrow p(x)=\dfrac{p}{s^d\Gamma(d/p)}x^{d-1}\exp\left(-(x/s)^p\right),
\end{array}
\ee
supported on $x\in(0,\infty)$. This defines the generalised $\gamma$ distribution, most readily implemented with a numeric CDF and QF. The mean of this distribution is $\mu_1=s\Gamma((d+1)/p)/\Gamma(d/p)$ and is removed before the fields are generated, while the field variable is $\tilde{x}=x/s$. The $\chi$, $\chi^2$, Weibull, Gamma, exponential and Erlang distributions are significant subsets of this generalised case.

This does not contain the normal or Laplace distributions, whose CDFs contain the error function rather than the incomplete Gamma. In these cases the PDF can be straightforwardly generalised to
\be
p(x)=\frac{p}{2s\Gamma(1/p)}\exp\left(-\left(|x|/s\right)^p\right),
\ee
supported on on $x\in(-\infty,\infty)$, which contains the Laplace distribution when $p=1$, the Gaussian distribution when $p=2$ and tends towards the uniform distribution when $p\rightarrow\infty$. The CDF of this distribution is
\be
C(x)=\frac{1}{2}\left(1+\mathrm{sgn}(x)P\left(1/p,\left(|x|/s\right)^p\right)\right) .
\ee
The mean is zero, and fields are generated with unit scale $s$.

Support for these distributions is included in the code; where the subcases have been separately considered the explicit forms are used rather than the above, which removes a small but potential cause of numerical noise (and due to the implementation allows for easier control of output directories). Naturally other generalisations of the gamma-type and error-type distributions are possible and can be readily included in the code.

\subsection{Fully Numeric Approaches}
\label{NumericDistributions}
\noindent Sometimes it is either not possible or inconvenient to specify a PDF analytically. In these cases, the user can specify a PDF numerically, which should be at as high a resolution as possible. The mean and standard deviation is calculated, the field rescaled appropriately and the PDF normalised, and the CDF and QF evaluated numerically. This approach can be demonstrated firstly with a distribution that is fully specified -- such as the uniform distribution -- and secondly with an arbitrarily intangible PDF.

\begin{figure*}
\includegraphics[width=\textwidth]{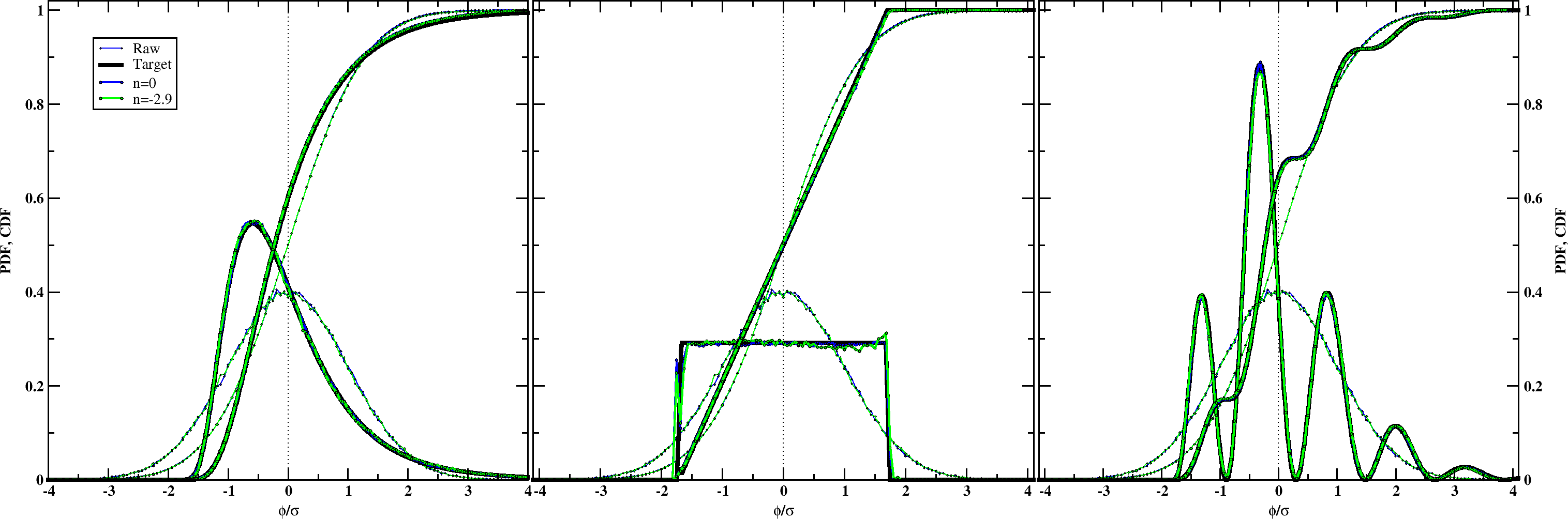}
\caption{The PDFs and CDFs for the log-normal distribution with $s=1/2$ (left), for a uniform distribution specified entirely numerically (middle), and for the distribution in equation (\ref{ArbitraryPDF}) (right).}
\label{Numerics}
\end{figure*}

\vspace{0.5em}\noindent\textbf{Uniform Distributions}\newline
\noindent Consider the PDF
\be
p(x)=\frac{1}{5}, \quad x\in (0,5)
\ee
and vanishing otherwise. An input array is primed with this PDF and passed into the code, which evaluates $\mu_1$ and $\sigma$ by direct integration; these values are retained in case the user wishes to reintroduce them. The mean is removed through a translation along the field axis and the field units changed to those of the standard deviation. The PDF is then normalised and the CDF and QF calculated. The resulting statistics are presented in Figure \ref{Numerics} and it is evident that for this extremely simple example the implementation is robust.

\vspace{0.5em}\noindent\textbf{Arbitrary PDF}
A more challenging test is to define the rather whimsical PDF
\be
p(x)=\frac{(1+\alpha^2)^3x^2\sin(\alpha x)\exp(-x)}{2(1+3\alpha+3\alpha^2-\alpha^3+3\alpha^4+\alpha^6)}
\ee
on the positive real line and with $\alpha>0$. Exact results for the mean and standard deviation can be found but are not very instructive. For concrete results take $\alpha=\pi$, implying
\be
\label{ArbitraryPDF}
\begin{array}{c}
p(x)=0.508546154x^2\left(1+\sin(\pi x)\right)\exp(-x), \\
\mu_1=3.026913369, \quad \sigma=1.744660857 .
\end{array}
\ee
The same procedure is followed as before and the resulting statistics are presented in the right panel of Figure \ref{Numerics}.

\section{Discussion}
\noindent I have presented an implementation of a spectral distortion/quantile transformation algorithm based on those of \cite{Grigoriu:1984,Yamazaki:1988,Grigoriu:1998,Deodatis:2001,Masters:2003}, similar to an implementation presented in \cite{Vio:2001cm,Vio:2002zr}. This new implementation is highly modular and has been constructed to allow a user to pass an arbitrary PDF, either defined in the calling routine or defined in a file stored on the drive. The algorithm generates a Gaussian field with a specific input power spectrum and passes it through a quantile transform, resulting in a field with the desired statistical nature and the desired power spectrum. The statistics are at the level of the probability distribution function (or, equivalently, the cumulative distribution function) and therefore do not rely on particular configurations of a bi- or a trispectrum. The input power spectrum need be found only once for a specified target PDF, spectrum and grid and can then be stored for future use with the saving of a significant amount of processing time. I have demonstrated the success of the code employing small grids and a wide range of probability distribution functions. The code is available at {\tt http://sourceforge.net/projects/nongaussian}, and serves as complementary to CMB-oriented codes such as presented in \cite{Rocha:2004ke,Fergusson:2009nv}.

Such a code has a wide applicability including areas from analysis of the CMB non-Gaussianity employing alternative probes to the bi- and trispectra, in generating initial conditions for $n$-body codes and smaller-scale simulations of cluster, galactic and molecular cloud dynamics, to studying the non-Gaussianity of the LSS. In many of these situations it cannot necessarily be assumed that the fields involved are Gaussian in nature, or have power spectra compatible with the manipulation of Gaussian fields. Using this code it is possible, for instance, for data analysts to consider arbitrary tests of CMB and LSS non-Gausianity without being constrained in their choises of non-Gaussian form, or encountering limits on the dynamic range from aliasing when taking products of Gaussian fields. Far from implying that the study of non-Gaussianities is not worthwhile, the tight Planck constraints provide a powerful dataset to tightly constrain mechanisms that produce non-Gaussianities in cosmology, from primordial signatures generated at an inflationary epoch, to nonlinear physics in the later universe and, as such, provide a powerful test of realistic models of cosmology. To fully exploit the rich data available in this field we have to be able to fully control both the statistics and the correlation structure of statistical realisations; this code will help towards that goal.

\begin{acknowledgments}
\noindent I wish to thank Yashar Akrami and Frode Hansen for discussions and support during the development of this work, and gratefully acknowledge routines I originally developed with Robert Crittenden which have been adapted and extended for this work. I am also grateful to Richard Brown for contributing I/O and FFT routines.
\end{acknowledgments}

\widetext{\bibliography{Paper}}

\begin{thebibliography}{39}
\expandafter\ifx\csname natexlab\endcsname\relax\def\natexlab#1{#1}\fi
\expandafter\ifx\csname bibnamefont\endcsname\relax
  \def\bibnamefont#1{#1}\fi
\expandafter\ifx\csname bibfnamefont\endcsname\relax
  \def\bibfnamefont#1{#1}\fi
\expandafter\ifx\csname citenamefont\endcsname\relax
  \def\citenamefont#1{#1}\fi
\expandafter\ifx\csname url\endcsname\relax
  \def\url#1{\texttt{#1}}\fi
\expandafter\ifx\csname urlprefix\endcsname\relax\def\urlprefix{URL }\fi
\providecommand{\bibinfo}[2]{#2}
\providecommand{\eprint}[2][]{\url{#2}}

\bibitem[{\citenamefont{Ade et~al.}(2013)}]{Ade:2013ydc}
\bibinfo{author}{\bibfnamefont{P.}~\bibnamefont{Ade}} \bibnamefont{et~al.}
  (\bibinfo{collaboration}{Planck Collaboration}) (\bibinfo{year}{2013}),
  \eprint{arXiv:1303.5084}.

\bibitem[{\citenamefont{Byrnes et~al.}(2010)\citenamefont{Byrnes, Gerstenlauer,
  Nurmi, Tasinato, and Wands}}]{Byrnes:2010ft}
\bibinfo{author}{\bibfnamefont{C.~T.} \bibnamefont{Byrnes}},
  \bibinfo{author}{\bibfnamefont{M.}~\bibnamefont{Gerstenlauer}},
  \bibinfo{author}{\bibfnamefont{S.}~\bibnamefont{Nurmi}},
  \bibinfo{author}{\bibfnamefont{G.}~\bibnamefont{Tasinato}}, \bibnamefont{and}
  \bibinfo{author}{\bibfnamefont{D.}~\bibnamefont{Wands}},
  \bibinfo{journal}{JCAP} \textbf{\bibinfo{volume}{1010}}, \bibinfo{pages}{004}
  (\bibinfo{year}{2010}), \eprint{arXiv:1007.4277}.

\bibitem[{\citenamefont{Mizuno et~al.}(2009)\citenamefont{Mizuno, Arroja, and
  Koyama}}]{Mizuno:2009mv}
\bibinfo{author}{\bibfnamefont{S.}~\bibnamefont{Mizuno}},
  \bibinfo{author}{\bibfnamefont{F.}~\bibnamefont{Arroja}}, \bibnamefont{and}
  \bibinfo{author}{\bibfnamefont{K.}~\bibnamefont{Koyama}},
  \bibinfo{journal}{Phys.Rev.} \textbf{\bibinfo{volume}{D80}},
  \bibinfo{pages}{083517} (\bibinfo{year}{2009}), \eprint{arXiv:0907.2439}.

\bibitem[{\citenamefont{Suyama et~al.}(2013)\citenamefont{Suyama, Takahashi,
  Yamaguchi, and Yokoyama}}]{Suyama:2013nva}
\bibinfo{author}{\bibfnamefont{T.}~\bibnamefont{Suyama}},
  \bibinfo{author}{\bibfnamefont{T.}~\bibnamefont{Takahashi}},
  \bibinfo{author}{\bibfnamefont{M.}~\bibnamefont{Yamaguchi}},
  \bibnamefont{and} \bibinfo{author}{\bibfnamefont{S.}~\bibnamefont{Yokoyama}}
  (\bibinfo{year}{2013}), \eprint{arXiv:1303.5374}.

\bibitem[{\citenamefont{Hansen et~al.}(2002)\citenamefont{Hansen, Marinucci,
  Natoli, and Vittorio}}]{Hansen:2002db}
\bibinfo{author}{\bibfnamefont{F.~K.} \bibnamefont{Hansen}},
  \bibinfo{author}{\bibfnamefont{D.}~\bibnamefont{Marinucci}},
  \bibinfo{author}{\bibfnamefont{P.}~\bibnamefont{Natoli}}, \bibnamefont{and}
  \bibinfo{author}{\bibfnamefont{N.}~\bibnamefont{Vittorio}},
  \bibinfo{journal}{Phys.Rev.} \textbf{\bibinfo{volume}{D66}},
  \bibinfo{pages}{063006} (\bibinfo{year}{2002}),
  \eprint{arXiv:astro-ph/0206501}.

\bibitem[{\citenamefont{Hansen et~al.}(2003)\citenamefont{Hansen, Marinucci,
  and Vittorio}}]{Hansen:2003bu}
\bibinfo{author}{\bibfnamefont{F.~K.} \bibnamefont{Hansen}},
  \bibinfo{author}{\bibfnamefont{D.}~\bibnamefont{Marinucci}},
  \bibnamefont{and} \bibinfo{author}{\bibfnamefont{N.}~\bibnamefont{Vittorio}},
  \bibinfo{journal}{Phys.Rev.} \textbf{\bibinfo{volume}{D67}},
  \bibinfo{pages}{123004} (\bibinfo{year}{2003}),
  \eprint{arXiv:astro-ph/0302202}.

\bibitem[{\citenamefont{Ducout et~al.}(2012)\citenamefont{Ducout, Bouchet,
  Colombi, Pogosyan, and Prunet}}]{Ducout:2012it}
\bibinfo{author}{\bibfnamefont{A.}~\bibnamefont{Ducout}},
  \bibinfo{author}{\bibfnamefont{F.}~\bibnamefont{Bouchet}},
  \bibinfo{author}{\bibfnamefont{S.}~\bibnamefont{Colombi}},
  \bibinfo{author}{\bibfnamefont{D.}~\bibnamefont{Pogosyan}}, \bibnamefont{and}
  \bibinfo{author}{\bibfnamefont{S.}~\bibnamefont{Prunet}}
  (\bibinfo{year}{2012}), \eprint{arXiv:1209.1223}.

\bibitem[{\citenamefont{Hindmarsh et~al.}(2009)\citenamefont{Hindmarsh,
  Ringeval, and Suyama}}]{Hindmarsh:2009qk}
\bibinfo{author}{\bibfnamefont{M.}~\bibnamefont{Hindmarsh}},
  \bibinfo{author}{\bibfnamefont{C.}~\bibnamefont{Ringeval}}, \bibnamefont{and}
  \bibinfo{author}{\bibfnamefont{T.}~\bibnamefont{Suyama}},
  \bibinfo{journal}{Phys.Rev.} \textbf{\bibinfo{volume}{D80}},
  \bibinfo{pages}{083501} (\bibinfo{year}{2009}), \eprint{arXiv:0908.0432}.

\bibitem[{\citenamefont{Bruni et~al.}(2012)\citenamefont{Bruni, Crittenden,
  Koyama, Maartens, Pitrou et~al.}}]{Bruni:2011ta}
\bibinfo{author}{\bibfnamefont{M.}~\bibnamefont{Bruni}},
  \bibinfo{author}{\bibfnamefont{R.}~\bibnamefont{Crittenden}},
  \bibinfo{author}{\bibfnamefont{K.}~\bibnamefont{Koyama}},
  \bibinfo{author}{\bibfnamefont{R.}~\bibnamefont{Maartens}},
  \bibinfo{author}{\bibfnamefont{C.}~\bibnamefont{Pitrou}},
  \bibnamefont{et~al.}, \bibinfo{journal}{Phys. Rev.}
  \textbf{\bibinfo{volume}{D85}}, \bibinfo{pages}{041301}
  (\bibinfo{year}{2012}), \eprint{arXiv:1106.3999}.

\bibitem[{\citenamefont{{Bardeen} et~al.}(1986)\citenamefont{{Bardeen}, {Bond},
  {Kaiser}, and {Szalay}}}]{Bardeen:1986}
\bibinfo{author}{\bibfnamefont{J.~M.} \bibnamefont{{Bardeen}}},
  \bibinfo{author}{\bibfnamefont{J.~R.} \bibnamefont{{Bond}}},
  \bibinfo{author}{\bibfnamefont{N.}~\bibnamefont{{Kaiser}}}, \bibnamefont{and}
  \bibinfo{author}{\bibfnamefont{A.~S.} \bibnamefont{{Szalay}}},
  \bibinfo{journal}{Astrophys. J.} \textbf{\bibinfo{volume}{304}},
  \bibinfo{pages}{15} (\bibinfo{year}{1986}).

\bibitem[{\citenamefont{Sirko}(2005)}]{Sirko:2005uz}
\bibinfo{author}{\bibfnamefont{E.}~\bibnamefont{Sirko}},
  \bibinfo{journal}{Astrophys. J.} \textbf{\bibinfo{volume}{634}},
  \bibinfo{pages}{728} (\bibinfo{year}{2005}), \eprint{arXiv:astro-ph/0503106}.

\bibitem[{\citenamefont{Laureijs et~al.}(2011)\citenamefont{Laureijs, Amiaux,
  Arduini, Augueres, Brinchmann et~al.}}]{Laureijs:2011mu}
\bibinfo{author}{\bibfnamefont{R.}~\bibnamefont{Laureijs}},
  \bibinfo{author}{\bibfnamefont{J.}~\bibnamefont{Amiaux}},
  \bibinfo{author}{\bibfnamefont{S.}~\bibnamefont{Arduini}},
  \bibinfo{author}{\bibfnamefont{J.-L.} \bibnamefont{Augueres}},
  \bibinfo{author}{\bibfnamefont{J.}~\bibnamefont{Brinchmann}},
  \bibnamefont{et~al.} (\bibinfo{year}{2011}), \bibinfo{note}{{\tt
  http://sci.esa.int/science-e/www/area/index.cfm?fareaid=102}},
  \eprint{arXiv:1110.3193}.

\bibitem[{\citenamefont{{Taylor}}(2013)}]{2013IAUS..291..337T}
\bibinfo{author}{\bibfnamefont{A.~R.} \bibnamefont{{Taylor}}}, in
  \emph{\bibinfo{booktitle}{IAU Symposium}} (\bibinfo{year}{2013}), vol.
  \bibinfo{volume}{291} of \emph{\bibinfo{series}{IAU Symposium}}, pp.
  \bibinfo{pages}{337--341}, \bibinfo{note}{{\tt http://www.skatelescope.org}}.

\bibitem[{\citenamefont{{Weinberg} and {Cole}}(1992)}]{Weinberg:1992}
\bibinfo{author}{\bibfnamefont{D.~H.} \bibnamefont{{Weinberg}}}
  \bibnamefont{and} \bibinfo{author}{\bibfnamefont{S.}~\bibnamefont{{Cole}}},
  \bibinfo{journal}{Month. Not. Royal Astron. Soc.}
  \textbf{\bibinfo{volume}{259}}, \bibinfo{pages}{652} (\bibinfo{year}{1992}).

\bibitem[{\citenamefont{Contaldi and Magueijo}(2001)}]{Contaldi:2001wr}
\bibinfo{author}{\bibfnamefont{C.~R.} \bibnamefont{Contaldi}} \bibnamefont{and}
  \bibinfo{author}{\bibfnamefont{J.}~\bibnamefont{Magueijo}},
  \bibinfo{journal}{Phys. Rev.} \textbf{\bibinfo{volume}{D63}},
  \bibinfo{pages}{103512} (\bibinfo{year}{2001}),
  \eprint{arXiv:astro-ph/0101512}.

\bibitem[{\citenamefont{Vio et~al.}(2001)\citenamefont{Vio, Andreani, and
  Wamsteker}}]{Vio:2001cm}
\bibinfo{author}{\bibfnamefont{R.}~\bibnamefont{Vio}},
  \bibinfo{author}{\bibfnamefont{P.}~\bibnamefont{Andreani}}, \bibnamefont{and}
  \bibinfo{author}{\bibfnamefont{W.}~\bibnamefont{Wamsteker}}
  (\bibinfo{year}{2001}), \eprint{arXiv:astro-ph/0105107}.

\bibitem[{\citenamefont{Avila-Reese et~al.}(2003)\citenamefont{Avila-Reese,
  Colin, Piccinelli, and Firmani}}]{AvilaReese:2003ue}
\bibinfo{author}{\bibfnamefont{V.}~\bibnamefont{Avila-Reese}},
  \bibinfo{author}{\bibfnamefont{P.}~\bibnamefont{Colin}},
  \bibinfo{author}{\bibfnamefont{G.}~\bibnamefont{Piccinelli}},
  \bibnamefont{and} \bibinfo{author}{\bibfnamefont{C.}~\bibnamefont{Firmani}},
  \bibinfo{journal}{Astrophys. J.} \textbf{\bibinfo{volume}{598}},
  \bibinfo{pages}{36} (\bibinfo{year}{2003}), \eprint{arXiv:astro-ph/0306293}.

\bibitem[{\citenamefont{Rocha et~al.}(2005)\citenamefont{Rocha, Hobson, Smith,
  Ferreira, and Challinor}}]{Rocha:2004ke}
\bibinfo{author}{\bibfnamefont{G.}~\bibnamefont{Rocha}},
  \bibinfo{author}{\bibfnamefont{M.}~\bibnamefont{Hobson}},
  \bibinfo{author}{\bibfnamefont{S.}~\bibnamefont{Smith}},
  \bibinfo{author}{\bibfnamefont{P.}~\bibnamefont{Ferreira}}, \bibnamefont{and}
  \bibinfo{author}{\bibfnamefont{A.}~\bibnamefont{Challinor}},
  \bibinfo{journal}{Month. Not. Royal Astron. Soc.}
  \textbf{\bibinfo{volume}{357}}, \bibinfo{pages}{1} (\bibinfo{year}{2005}),
  \eprint{arXiv:astro-ph/0406136}.

\bibitem[{\citenamefont{Smith and Zaldarriaga}(2011)}]{Smith:2006ud}
\bibinfo{author}{\bibfnamefont{K.~M.} \bibnamefont{Smith}} \bibnamefont{and}
  \bibinfo{author}{\bibfnamefont{M.}~\bibnamefont{Zaldarriaga}},
  \bibinfo{journal}{Mon.Not.Roy.Astron.Soc.} \textbf{\bibinfo{volume}{417}},
  \bibinfo{pages}{2} (\bibinfo{year}{2011}), \eprint{arXiv:astro-ph/0612571}.

\bibitem[{\citenamefont{Fergusson et~al.}(2010)\citenamefont{Fergusson,
  Liguori, and Shellard}}]{Fergusson:2009nv}
\bibinfo{author}{\bibfnamefont{J.}~\bibnamefont{Fergusson}},
  \bibinfo{author}{\bibfnamefont{M.}~\bibnamefont{Liguori}}, \bibnamefont{and}
  \bibinfo{author}{\bibfnamefont{E.}~\bibnamefont{Shellard}},
  \bibinfo{journal}{Phys.Rev.} \textbf{\bibinfo{volume}{D82}},
  \bibinfo{pages}{023502} (\bibinfo{year}{2010}), \eprint{arXiv:0912.5516}.

\bibitem[{\citenamefont{Viel et~al.}(2009)\citenamefont{Viel, Branchini, Dolag,
  Grossi, Matarrese et~al.}}]{Viel:2008jj}
\bibinfo{author}{\bibfnamefont{M.}~\bibnamefont{Viel}},
  \bibinfo{author}{\bibfnamefont{E.}~\bibnamefont{Branchini}},
  \bibinfo{author}{\bibfnamefont{K.}~\bibnamefont{Dolag}},
  \bibinfo{author}{\bibfnamefont{M.}~\bibnamefont{Grossi}},
  \bibinfo{author}{\bibfnamefont{S.}~\bibnamefont{Matarrese}},
  \bibnamefont{et~al.}, \bibinfo{journal}{Month. Not. Royal Astron. Soc.}
  \textbf{\bibinfo{volume}{393}}, \bibinfo{pages}{774} (\bibinfo{year}{2009}),
  \eprint{arXiv:0811.2223}.

\bibitem[{\citenamefont{Scoccimarro}(2000)}]{Scoccimarro:2000qg}
\bibinfo{author}{\bibfnamefont{R.}~\bibnamefont{Scoccimarro}}
  (\bibinfo{year}{2000}), \eprint{arXiv:astro-ph/0002037}.

\bibitem[{\citenamefont{Liguori et~al.}(2003)\citenamefont{Liguori, Matarrese,
  and Moscardini}}]{Liguori:2003mb}
\bibinfo{author}{\bibfnamefont{M.}~\bibnamefont{Liguori}},
  \bibinfo{author}{\bibfnamefont{S.}~\bibnamefont{Matarrese}},
  \bibnamefont{and}
  \bibinfo{author}{\bibfnamefont{L.}~\bibnamefont{Moscardini}},
  \bibinfo{journal}{Astrophys. J.} \textbf{\bibinfo{volume}{597}},
  \bibinfo{pages}{57} (\bibinfo{year}{2003}), \eprint{arXiv:astro-ph/0306248}.

\bibitem[{\citenamefont{Grigoriu}(1984)}]{Grigoriu:1984}
\bibinfo{author}{\bibfnamefont{M.}~\bibnamefont{Grigoriu}},
  \bibinfo{journal}{J. Eng. Mech.} \textbf{\bibinfo{volume}{110}},
  \bibinfo{pages}{610} (\bibinfo{year}{1984}).

\bibitem[{\citenamefont{Yamazaki and Shinozuka}(1988)}]{Yamazaki:1988}
\bibinfo{author}{\bibfnamefont{F.}~\bibnamefont{Yamazaki}} \bibnamefont{and}
  \bibinfo{author}{\bibfnamefont{M.}~\bibnamefont{Shinozuka}},
  \bibinfo{journal}{J. Eng. Mech.} \textbf{\bibinfo{volume}{114}},
  \bibinfo{pages}{1183} (\bibinfo{year}{1988}).

\bibitem[{\citenamefont{Deodatis}(1996)}]{Deodatis:1996}
\bibinfo{author}{\bibfnamefont{G.}~\bibnamefont{Deodatis}},
  \bibinfo{journal}{J. Eng. Mech.} \textbf{\bibinfo{volume}{122}},
  \bibinfo{pages}{778} (\bibinfo{year}{1996}).

\bibitem[{\citenamefont{Grigoriu}(1998)}]{Grigoriu:1998}
\bibinfo{author}{\bibfnamefont{M.}~\bibnamefont{Grigoriu}},
  \bibinfo{journal}{J. Eng. Mech.} \textbf{\bibinfo{volume}{124}},
  \bibinfo{pages}{121} (\bibinfo{year}{1998}).

\bibitem[{\citenamefont{Gurley and Kareem}(1998)}]{Gurley:1998}
\bibinfo{author}{\bibfnamefont{K.~R.} \bibnamefont{Gurley}} \bibnamefont{and}
  \bibinfo{author}{\bibfnamefont{A.}~\bibnamefont{Kareem}},
  \bibinfo{journal}{Meccanica} \textbf{\bibinfo{volume}{33}},
  \bibinfo{pages}{309} (\bibinfo{year}{1998}).

\bibitem[{\citenamefont{Deodatis and Micaletti}(2001)}]{Deodatis:2001}
\bibinfo{author}{\bibfnamefont{G.}~\bibnamefont{Deodatis}} \bibnamefont{and}
  \bibinfo{author}{\bibfnamefont{R.~C.} \bibnamefont{Micaletti}},
  \bibinfo{journal}{J. Eng. Mech.} \textbf{\bibinfo{volume}{127}},
  \bibinfo{pages}{1284} (\bibinfo{year}{2001}).

\bibitem[{\citenamefont{Masters and Gurley}(2003)}]{Masters:2003}
\bibinfo{author}{\bibfnamefont{F.}~\bibnamefont{Masters}} \bibnamefont{and}
  \bibinfo{author}{\bibfnamefont{K.~R.} \bibnamefont{Gurley}},
  \bibinfo{journal}{J. Eng. Mech.} \textbf{\bibinfo{volume}{129}},
  \bibinfo{pages}{1418} (\bibinfo{year}{2003}).

\bibitem[{\citenamefont{Phoon et~al.}(2005)\citenamefont{Phoon, Huang, and
  Quek}}]{Phoon:2005}
\bibinfo{author}{\bibfnamefont{K.~K.} \bibnamefont{Phoon}},
  \bibinfo{author}{\bibfnamefont{H.~W.} \bibnamefont{Huang}}, \bibnamefont{and}
  \bibinfo{author}{\bibfnamefont{S.~T.} \bibnamefont{Quek}},
  \bibinfo{journal}{Prob. Eng. Mech.} \textbf{\bibinfo{volume}{20}}
  (\bibinfo{year}{2005}).

\bibitem[{\citenamefont{Bocchini and Deodatis}(2008)}]{Bocchini:2008}
\bibinfo{author}{\bibfnamefont{P.}~\bibnamefont{Bocchini}} \bibnamefont{and}
  \bibinfo{author}{\bibfnamefont{G.}~\bibnamefont{Deodatis}},
  \bibinfo{journal}{Prob. Eng. Mech.} \textbf{\bibinfo{volume}{23}}
  (\bibinfo{year}{2008}).

\bibitem[{\citenamefont{Shields et~al.}(2011)\citenamefont{Shields, Deodatis,
  and Bocchini}}]{Shields:2011}
\bibinfo{author}{\bibfnamefont{M.~D.} \bibnamefont{Shields}},
  \bibinfo{author}{\bibfnamefont{G.}~\bibnamefont{Deodatis}}, \bibnamefont{and}
  \bibinfo{author}{\bibfnamefont{P.}~\bibnamefont{Bocchini}},
  \bibinfo{journal}{Prob. Eng. Mech.} \textbf{\bibinfo{volume}{26}},
  \bibinfo{pages}{511} (\bibinfo{year}{2011}).

\bibitem[{\citenamefont{Sakamoto and
  Ghanem}(2002{\natexlab{a}})}]{Sakamoto:2002J}
\bibinfo{author}{\bibfnamefont{S.}~\bibnamefont{Sakamoto}} \bibnamefont{and}
  \bibinfo{author}{\bibfnamefont{R.}~\bibnamefont{Ghanem}},
  \bibinfo{journal}{J. Eng. Mech.} \textbf{\bibinfo{volume}{128}}
  (\bibinfo{year}{2002}{\natexlab{a}}).

\bibitem[{\citenamefont{Sakamoto and
  Ghanem}(2002{\natexlab{b}})}]{Sakamoto:2002P}
\bibinfo{author}{\bibfnamefont{S.}~\bibnamefont{Sakamoto}} \bibnamefont{and}
  \bibinfo{author}{\bibfnamefont{R.}~\bibnamefont{Ghanem}},
  \bibinfo{journal}{Prob. Eng. Mech.} \textbf{\bibinfo{volume}{17}}
  (\bibinfo{year}{2002}{\natexlab{b}}).

\bibitem[{\citenamefont{Vio et~al.}(2002)\citenamefont{Vio, Andreani, Tenorio,
  and Wamsteker}}]{Vio:2002zr}
\bibinfo{author}{\bibfnamefont{R.}~\bibnamefont{Vio}},
  \bibinfo{author}{\bibfnamefont{P.}~\bibnamefont{Andreani}},
  \bibinfo{author}{\bibfnamefont{L.}~\bibnamefont{Tenorio}}, \bibnamefont{and}
  \bibinfo{author}{\bibfnamefont{W.}~\bibnamefont{Wamsteker}},
  \bibinfo{journal}{Publ. Astron. Soc. Pac.}  (\bibinfo{year}{2002}),
  \eprint{arXiv:astro-ph/0207311}.

\bibitem[{\citenamefont{Pullen and Kamionkowski}(2007)}]{Pullen:2007tu}
\bibinfo{author}{\bibfnamefont{A.~R.} \bibnamefont{Pullen}} \bibnamefont{and}
  \bibinfo{author}{\bibfnamefont{M.}~\bibnamefont{Kamionkowski}},
  \bibinfo{journal}{Phys.Rev.} \textbf{\bibinfo{volume}{D76}},
  \bibinfo{pages}{103529} (\bibinfo{year}{2007}), \eprint{arXiv:0709.1144}.

\bibitem[{\citenamefont{Brown and Crittenden}(2005)}]{Brown:2005kr}
\bibinfo{author}{\bibfnamefont{I.}~\bibnamefont{Brown}} \bibnamefont{and}
  \bibinfo{author}{\bibfnamefont{R.}~\bibnamefont{Crittenden}},
  \bibinfo{journal}{Phys.Rev.} \textbf{\bibinfo{volume}{D72}},
  \bibinfo{pages}{063002} (\bibinfo{year}{2005}),
  \eprint{arXiv:astro-ph/0506570}.

\bibitem[{\citenamefont{Paoletti et~al.}(2009)\citenamefont{Paoletti, Finelli,
  and Paci}}]{Paoletti:2008ck}
\bibinfo{author}{\bibfnamefont{D.}~\bibnamefont{Paoletti}},
  \bibinfo{author}{\bibfnamefont{F.}~\bibnamefont{Finelli}}, \bibnamefont{and}
  \bibinfo{author}{\bibfnamefont{F.}~\bibnamefont{Paci}},
  \bibinfo{journal}{Month. Not. Royal Astron. Soc.}
  \textbf{\bibinfo{volume}{396}}, \bibinfo{pages}{523} (\bibinfo{year}{2009}),
  \eprint{arXiv:0811.0230}.

\end{thebibliography}

\end{document}